\documentclass[prd,aps,preprint,eqsecnum,floats,showpacs]{revtex4}

\begin{document}
\title{The restriction on the strong coupling constant in the IR region from
the 1D-1P splitting in bottomonium}
\author{A.M. Badalian and A.I. Veselov}
\affiliation{Institute of Theoretical and Experimental Physics,
B.Cheremushkinskaya 25, 117218 Moscow, Russia, }
\author{B.L.G. Bakker}
\affiliation{Vrije Universiteit, Department of Physics,
Amsterdam, The Netherlands}
\def\la{\mathrel{\mathpalette\fun <}}
\def\ga{\mathrel{\mathpalette\fun >}}
\def\fun#1#2{\lower3.6pt\vbox{\baselineskip0pt\lineskip.9pt
\ialign{$\mathsurround=0pt#1\hfil ##\hfil$\crcr#2\crcr\sim\crcr}}}
\newcommand{\vep}{\mbox{\boldmath${\it p}$}}
\newcommand{\veL}{\mbox{\boldmath${\it L}$}}
\newcommand{\lan}{\langle}
\newcommand{\ran}{\rangle}

\begin{abstract}
The $b\bar b$ spectrum is calculated with the use of a relativistic
Hamiltonian where the gluon-exchange potential between a quark and an
antiquark is taken as in background perturbation theory. We observed
that the splittings $\Delta_1=  \Upsilon({\rm 1D})-\chi_b({\rm 1P})$
and other splittings between low-lying states are very sensitive to the
QCD constant $\Lambda_{\rm V}(n_{\rm f})$ which occurs in the Vector
scheme, and good agreement with the experimental data is obtained for
$\Lambda_{\rm V}(2$-loop, $n_{\rm f}=5)= 325\pm 10$ MeV which
corresponds to the conventional $\Lambda_{\overline{\rm MS}} (2-$loop,
$n_{\rm f}=5)= 238\pm 7$ MeV, $\alpha_s(2-$loop, $M_Z)=0.1189\pm
0.0005,$ and to a large freezing value of the background coupling:
$\alpha_{\rm crit} (2$-loop, $q^2=0)=\alpha_{\rm crit} (2$-loop, $r\to
\infty)=0.58\pm 0.02$. If the asymptotic freedom behavior of the
coupling is neglected and an effective freezing coupling $\alpha_{\rm
static}=const$ is introduced, as in the Cornell potential,  then
precise agreement with $\Delta_1({\rm exp})$ and $\Delta_2({\rm exp})$
can be reached for the rather large Coulomb constant $\alpha_{\rm
static} =0.43\pm 0.02$.  We predict a value for the mass $M({\rm 2D}) =
10451\pm 2$ MeV.
\end{abstract}

\pacs{11.15.Tk, 12.38.Lg, 14.40.Gx}
\maketitle

\section{Introduction}
\label{sec.01}

Recently the CLEO collaboration has discovered the first stable D-wave
state in bottomonium with a mass $M({\rm 1D})=10162.2\pm 1.6$ MeV which
is consistent with the $J=2$ assignment \cite{ref.01} The observation
of this state is also a great success of the potential model (PM) since
just in the framework of the PM approach the strategy how to observe
the 1D state was developed and correct values for the radiative
transitions and the $M({\rm 1D})$ mass were obtained
\cite{ref.02,ref.03}. A study of the orbital and radial $b\bar{b}$
excitations which have relatively large sizes is also very important
for a better understanding of the fundamental interaction in the
infrared (IR) region. At present different conceptions about the value
of the strong coupling in the IR region exist. In lattice QCD one finds
for the coupling constant in the static potential, parametrized as a
linear plus Coulomb potential, the small value $\alpha_{\rm static} =
0.23\, (n_f = 0) - 0.30\, (n_f = 3)$ \cite{ref.04}. In phenomenological
potentials the Coulomb constant is larger but spreads over a wide range
from $\alpha_{\rm static}\approx 0.33-0.39$ \cite{ref.05,ref.06} up to
the rather large values $0.42-0.45$ adopted in recent calculations
\cite{ref.07,ref.08}. If the asymptotic freedom (AF) behaviour is taken
into account, then an even larger freezing value of the physical
coupling, $\alpha_s (1\,{\rm GeV})\sim 0.9 \pm 0.1$, was determined
from the analysis of the hadronic decays of the $\tau$-lepton
\cite{ref.09}. Our point of view discussed here is that after the
discovery of the $\Upsilon({\rm 1D})$ the bottomonium spectrum already
contains enough information to determine the freezing value of the
coupling constant in an unambiguous way.

Among the excited states in bottomonium the most precise information
about the static interaction  at large $r$ can be extracted from an
analysis of the 1P, 2P, 1D, and 2D states, because they have
large sizes,  lie below the open flavor threshold, and have no hadronic
shifts. We shall show that instead of the absolute masses it is more
convenient to use  the splittings between $\underline{\rm orbital}$
excitations since they are less sensitive to the choice of such
parameters as the $b$-quark mass $m$ and in the  mass differences the
relativistic corrections (about $\la 20$ MeV) are partially or
completely cancelled.
Note that for these states the spin-averaged masses $M_{\rm cog}(nL)$,
and therefore also the splittings between them, are known with great
accuracy, $\sim 1$ MeV. For the nS states (since the states
$\eta_b(nS)$ are  still unobserved) the accuracy of the spin-averaged
masses $M_{\rm cog}(nS)$ as well as the splittings 2S-1S and 1P-1S is
only about 10 MeV.

From experiment we take two splittings between the spin-averaged
masses for orbital excitations:
\begin{eqnarray}
 \Delta_1 & =  & M_{\rm cog}({\rm 1D}) - M_{\rm cog}({\rm 1P}) \; =
 \; 262.1\pm
 2.2\;({\rm exp})^{+ 1}_{- 0}\; ({\rm th}) ~\mbox{MeV}, 
\nonumber \\
 \Delta_2 & = & M_{\rm cog}({\rm 2P})-M_{\rm cog}({\rm 1P}) \; =
 \; 360.1\pm1.2\;({\rm exp})~\mbox{MeV},
\label{eq.1.01}
\end{eqnarray}
which are measured, as well as the masses $M_{\rm cog}({\rm
1P})=9900.1\pm 0.6$ MeV, $M_{\rm cog}({\rm 2P})=10260.0\pm 0.6$ MeV,
and $M({\rm 1D}_2)= 10162.2\pm 1.6$ MeV with high accuracy. However,
for $M_{\rm cog}({\rm 1D})$ one needs to take into account the very
small difference between $M_{\rm cog}({\rm 1D})$ and $M({\rm 1D}_2)$
coming from the fine structure splittings: $M_{\rm cog}({\rm
1D})=M_{\rm exp}({\rm 1D}_2) + \delta_{\rm FS}$.  In Ref.~\cite{ref.03}
$\delta_{\rm FS}$ was found to be small lying in the range $0\leq
\delta_{\rm FS} \le 1$ MeV  so that $\delta_{\rm FS}$  was included in
the theoretical error  in Eq.~(\ref{eq.1.01}).

In this paper we show that the splitting $\Delta_1$ weakly depends on
the kinematics, the string tension $\sigma$ and the pole mass of the
$b$ quark, being at the same time very sensitive to the gluon-exchange
potential used.  Therefore $\Delta_1$ (as well as $\Delta_3=M({\rm 2D})
- M({\rm 2P}))$ can be used as  a probe of the freezing value of the
strong coupling in the IR region.

\section{Hamiltonian for spinless quark and antiquark}
\label{sec.02}

In  calculations of the $b\bar b$ spectrum we use  the relativistic
(string) Hamiltonian $H_{\rm R}$ which is derived from the
Fock-Feynman-Schwinger (FFS) representation of the meson Green's
function in QCD \cite{ref.10,ref.11,ref.12}. This Hamiltonian and the
calculated meson masses $M(nL)$ contain only fundamental quantities:
the current (pole) mass of the $b$ quark, the string tension $\sigma$,
and the QCD constant $\Lambda_{\overline{\rm MS}} (n_{\rm f})$ . It is
known that in QCD, besides the current quark masses, only one
``external'' mass scale must occur, say the QCD constant
$\Lambda_{\overline{\rm MS}}$, while the string tension has to be
expressed through this scale. Such a relation is established in Lattice
QCD, however, in analytical approaches, including BPT, a relation
between $\sigma$ and $\Lambda$ is still not derived.

The Hamiltonian $H_{\rm R}$ is derived using several approximations.\\
- First, the $b\bar b$ states below the open beauty threshold are described
in the one-channel approximation, i.e., the creation of an additional
quark-antiquark pair is neglected.\\
- Second, the spin-dependent terms in the $Q\bar Q$ interaction are
considered as a perturbation.\\
- Third, the minimal area law is  used for the vacuum average of the
Wilson loop. Then the Hamiltonian $H_{\rm R}$ has the following simple form:
\begin{equation}
 H_{\rm R} =\omega+ \frac{m^2_q}{\omega} +\frac{\vep^2}{\omega}+V_{\rm st}(r)
\label{eq.2.01}
\end{equation}
where $\vep^2=p^2_r+ \veL^2/r^2$ and the ``einbein'' variable $\omega$
appears when in the Green's function in the FFS representation one goes
over from the proper time $\tau$  to the actual time $t$: $2\omega =
\frac{dt}{d\tau}$. According to the quantization procedure the
operator $\omega$ is determined from the extremum condition
\cite{ref.10,ref.13}:
\begin{equation}
\frac{\partial H_{\rm R}}{\partial \omega} = 0\Longrightarrow \omega =
\sqrt{\vep^2+m^2_q}.
\label{eq.2.02} 
\end{equation}
Therefore the Hamiltonian (\ref{eq.2.01}) reduces to an expression which does
not explicitly depend on $\omega$,
\begin{equation}
 H_{\rm R}=2\sqrt{\vep^2+m^2_q} + V_{\rm st}(r).
\label{eq.2.03} 
\end{equation}
The kinetic term in Eq.~(\ref{eq.2.03}) was postulated many years ago
\cite{ref.14} and successfully used in the relativized potential model
\cite{ref.06,ref.07,ref.13,ref.14,ref.15,ref.16}, however, the derivation of
this term directly from the QCD meson Green's function was   done for
the first time in Ref. \cite{ref.10}. The Hamiltonian (\ref{eq.2.01})
or (\ref{eq.2.03}) also implies a definite prescription for $m_q$ while
in the PM the quark mass $m_q$ is usually considered as a fitting
parameter.

Let us consider two limiting cases. The first one occurs when the
nonperturbative (NP) contribution dominates in the $Q\bar Q$ potential
(at relatively large distances) and the $Q\bar Q$ gluon-exchange term
for excited states can be considered as a perturbation \cite{ref.13}.
Such an approximation is valid for light mesons $(m_q=0)$. Then by
derivation the mass $m_q$ in Hamiltonian (\ref{eq.2.01}) coincides with
the Lagrangian current mass (the difference between current and pole
masses can be neglected in the approximation considered):  
\begin{equation}
 m_q=\bar m_q(\bar m_q),\quad V(r)= V_{\rm st}^{\rm NP}(r) =\sigma r.
\label{eq.2.04} 
\end{equation} 
One must also take into account the NP self-energy correction to the
quark masses \cite{ref.17}, however, in bottomonium the NP self-energy
contribution to $H_{\rm R}$ appears to be compatible with zero ($\sim 3$ MeV)
and can be neglected \cite{ref.18}.  For the linear potential the splitting
$\Delta_1$ between the orbital excitations 1D and 1P turns out to be
$\sim 160\div 170$ MeV, i.e., much smaller then the experimental value
(\ref{eq.1.01}).

A different situation takes place at small distances where the 
perturbative $Q\bar Q$ interaction  dominates while in first
approximation the string interaction can be neglected. Then due to
perturbative self-energy corrections the pole mass $m_q$ appears
in the Hamiltonian (\ref{eq.2.01}) and for a consistent description
$m_q$(pole) has to be taken in the same $n$-loop approximation
as the perturbative $Q\bar Q$ potential:
\begin{equation}
 m_q= m_q^{\rm pole} ({\rm n-loop}),\quad V_{\rm st}^{\rm P} (r)
 =-\frac43\frac{\alpha_{\rm V}({\rm n-loop})}{r}.
\label{eq.2.05} 
\end{equation}

For low-lying $b\bar b$ states this type of calculations was performed
in Refs.~\cite{ref.19} (with the use also of the $1/m_q$ expansion of the
kinetic term in Eq.~(\ref{eq.2.03})).

The perturbative static potential in Eq.~(\ref{eq.2.05}) contains the
vector coupling $\alpha_{\rm V}(r)$ in coordinate space. Its relation to the
vector coupling $\alpha_{\rm V}(q)$ in momentum space and to the QCD strong
coupling constant $\alpha_s(q)$ in the $\overline{\rm MS}$ scheme (up to
three-loops) was studied in detail in Refs.~\cite{ref.20} where the constant
$\Lambda_{\rm V}(\Lambda_{\rm R})$, which determines the perturbative vector coupling
$\alpha_{\rm V}$ in momentum (coordinate) space, was expressed through  the
QCD constant $\Lambda_{\overline{\rm MS}}$:
\begin{eqnarray}
 \Lambda_{\rm V}(n_{\rm f}) & = & \Lambda_{\overline{\rm MS}}(n_{\rm f}) \exp
 \left(\frac{a_1}{2\beta_0}\right)
\nonumber \\
 \Lambda_{\rm R}(n_{\rm f}) & = & \Lambda_{\rm V}^{\gamma_E},
\label{eq.2.06}
\end{eqnarray}
where $a_1=\frac{31}{3}-\frac{10}{9} n_{\rm f}$,~~ $\beta_0=11-\frac23 n_{\rm f},$
and $\gamma_E$ is the Euler constant.

If one takes the conventional value 
$\Lambda_{\overline{\rm MS}}^{(5)}(2$-loop) = 216$\pm25$ MeV, which
corresponds to the ``world average'' $\alpha_s(2-$loop,
$M_Z)=0.1172\pm 0.0020$ \cite{ref.21}, then from the definitions
Eq.~(\ref{eq.2.06}) it follows that
\begin{eqnarray}
\Lambda_{\rm V}^{(5)} (2-{\rm loop}) & = & 295\pm 34\, {\rm MeV},
\nonumber \\
\Lambda_{\rm R}^{(5)} (2-{\rm loop}) & = & 525\pm 61\, {\rm MeV}.
\label{eq.2.07}
\end{eqnarray}
Thus  both $\Lambda_{\rm V}^{(5)}$ and $\Lambda_{\rm R}^{(5)}$ appear to
be essentially larger than $\Lambda^{(5)}_{\overline {\rm MS}}$ and
therefore the perturbative description (where $\Lambda_{\rm R} r\ll1$ or
$r\ll 0.3$ fm),  in a strict sense, can be applied only to the ground state
$\Upsilon ({\rm 1S})$. (The validity of the perturbative static
potential in quenched approximation was discussed in Refs.
\cite{ref.22,ref.23}.) Since  the orbital and radial excitations in
bottomonium have rather large sizes, e.g., $R({\rm 1P}) =0.4$ fm,
$R({\rm 1D})=0.5$ fm, $R({\rm 2P}) =0.6$ fm, $R({\rm 2D}) =0.7$ fm,
for them the $Q\bar Q$ gluon exchange cannot be described in the
framework of PQCD alone and a NP modification of $\alpha_s (\alpha_{\rm V})$ at
large distances (IR freezing) should be taken into account. In our
calculations we shall use $\Lambda_{\rm V}^{(5)}$ from the range (\ref{eq.2.07}).

\section{Static potential}
\label{sec.03}

To describe the $b\bar b$ spectrum as a whole we shall use here
the static $Q\bar Q$ potential as it is defined in background
perturbation theory (BPT), where the influence of the background
field on the gluon exchange is taken into account
\cite{ref.11,ref.24}:
\begin{equation}
 V_{\rm st} =\sigma r + V_{\rm B}(r),
\label{eq.3.01}
\end{equation}
\begin{equation}
 V_{\rm B}(r) =-\frac43 \frac{\alpha_{\rm B}(r)}{r}.
\label{eq.3.02}
\end{equation}
Here the background coupling $\alpha_{\rm B}(r) $ is defined through the
Fourier transform of the potential $V_{\rm B}(q) $ in momentum space
\cite{ref.23}; it gives
\begin{equation}
 \alpha_{\rm B} (r) =\frac{2}{\pi} \int^\infty_0 \frac{dq}{q} \sin (qr)
 \alpha_{\rm B}(q), 
\label{eq.3.03} 
\end{equation}
where the background coupling in momentum space $\alpha_{\rm B} (q)$ is
defined over the whole region $0\leq q < \infty$. For example, in
two-loop approximation
\begin{eqnarray}
 \alpha_{\rm B}(q) & =& \frac{4\pi}{\beta_0} \frac{1}{t_{\rm B}} \left
 (1-\frac{\beta_1}{\beta^2_0} \frac{\ln t_{\rm B}}{t_{\rm B}} \right),
\nonumber \\
 t_{\rm B} & =& \ln \frac{q^2+M^2_{\rm B}}{\Lambda^2_{\rm V}}, 
\label{eq.3.04}
\end{eqnarray}
where the background mass  $M_{\rm B}$  under the logarithm is determined by
the lowest hybrid excitation of the string. This mass cannot be
considered as an additional (fitting) parameter since $M_{\rm B}$ itself can
be calculated either on the lattice or with the use of the
corresponding Hamiltonian for a  hybrid \cite{ref.12}. We take here the
value of $M_{\rm B}$ which was obtained in Ref. \cite{ref.23} from  a fit to
the static potential on the lattice at small distances \cite{ref.22},
\begin{equation}
 M_{\rm B}= 1.00\pm 0.05\; {\rm GeV}.
\label{eq.3.05}
\end{equation}
Note that the most transparent way suggested to derive the expression
(\ref{eq.3.04}) for $\alpha_{\rm B}(q)$  is to consider a large number of
colors, $N_c\gg1$ \cite{ref.11}.

The background coupling in momentum space $\alpha_{\rm B}(q)$ has an
important feature--the correct PQCD limit at $q^2\gg M^2_{\rm B}$ (as
in Refs. \cite{ref.20}) and therefore the constant $\Lambda_{\rm V}$ in the 
Vector-scheme (\ref{eq.3.04}) used in our calculations, is determined by
the conventional $\Lambda_{\overline{\rm MS}} (n_{\rm f})$ according  to the
relation (\ref{eq.2.06}).

One should keep in mind that the additive form of the static potential
(\ref{eq.3.01}) is automatically obtained in the framework of BPT in
the lowest approximation when the interference terms are neglected.
This form is well supported by the lattice data \cite{ref.25} where the
static potential is taken as the sum of a Coulomb-type term plus linear
term at separations $r \ge 0.2$ fm. Note that both potentials satisfy the
Casimir scaling law with an accuracy of a few percent \cite{ref.26}.

In the next approximation of BPT the interference between
perturbative and NP effects appears in the form of the background
mass $M_{\rm B}$ (\ref{eq.3.04}) which ensures the IR freezing of
$\alpha_{\rm B}(q)$. One can see from Eq.~(\ref{eq.3.04}) that the only
effect of $M_{\rm B}$ is to moderate the IR behavior of the perturbative
potential, whereby the Landau ghost pole and IR renormalons
disappear, while the short distance behavior, as well as the
Casimir scaling property, stay intact \cite{ref.11}.
Note that the logarithm in the coupling (\ref{eq.3.04}) formally coincides
with that  suggested in Refs. \cite{ref.27} with $M^2_{\rm B}=4 m^2_g$ where
a picture with the gluon acquiring an effective mass $m_g$ was
suggested. However, since a physical gluon has no mass, the
parameter ($2m_g)^2$ needs to be reinterpreted in the correct way.

Our calculations of the $b\bar b$ spectrum are done for the number of
flavors $n_{\rm f}=5$ when the QCD constant $\Lambda_{\overline{\rm MS}}^{(5)}$
is well known from high energy processes,
$\Lambda_{\overline{\rm MS}}^{(5)} = 216\pm 25$ MeV, which corresponds to
$\alpha_s(M_z) = 0.117\pm 0.002$ \cite{ref.20} and correspondingly in
the Vector-scheme $\Lambda_{\rm V}^{(5)}= 295\pm 34$ MeV.  Later in our
calculations  we show that the splittings $\Delta_1$ and $\Delta_2$
appear to be in agreement with experiment only for those values of
$\Lambda_{\rm V}^{(5)}$(2-loop) which are close to the upper limit in
Eq.~(\ref{eq.2.07}),
\begin{equation}
 \Lambda_{\rm V}^{(5)} (2{\rm -loop})= 325\pm 10\; {\rm MeV},
\label{eq.3.06}
\end{equation}
which corresponds to
\begin{equation}
 \Lambda_{\overline{\rm MS}}^{(5)} ({\rm 2-loop}) =238\pm 7\; {\rm MeV},
 \quad \alpha_s(M_z) =0.1189\pm 0.0005.
\label{eq.3.07}
\end{equation}
From the definition (\ref{eq.3.04}) and for $\Lambda_{\rm V}^{(5)}$
(\ref{eq.3.06}) the freezing (critical) value of the  background
coupling $\alpha_{\rm B}(q=0)=\alpha_{\rm B}(r\to \infty)$ is expected to be
\begin{eqnarray}
 \alpha_{\rm crit} (n_{\rm f}=5) & =& 0.56\pm 0.01\; (M_{\rm B}=1.0\; {\rm GeV}),
\nonumber \\
 \alpha_{\rm crit} (n_{\rm f}=5) & =& 0.59\pm 0.01\; (M_{\rm B}=0.95\; {\rm GeV }). 
\label{eq.3.08} 
\end{eqnarray}
Notice that from Eq.~(\ref{eq.3.03}) $\alpha_{\rm crit}$ turns out to be
the same in momentum and coordinate space.

For a value of $ \Lambda_{\rm V}^{(5)}$ close to the lower bound the
freezing value $\alpha_{\rm crit}$ Eq.~(\ref{eq.2.07}) turns out to be
essentially smaller, e.g., for $\Lambda^{(5)}_{ \overline{\rm MS}} =190$
MeV $(\Lambda_{\rm V}^{(5)} =260$ MeV) the value $\alpha_{\rm crit} =0.46$ is
obtained. For such a choice of $\Lambda_{\rm V}^{(5)}$ agreement with
experiment for the $b\bar b$ spectrum cannot be reached.

For $\Lambda_{\overline{\rm MS}}^{(5)}$, Eq.~(\ref{eq.3.07}), the pole mass of
the $b$ quark can be  determined according to  the standard procedure
(as in PQCD) and for the conventional value $\bar{m}_b(\bar{m}_b)= 4.23\pm
0.03$ GeV one obtains in two-loop approximation
\begin{equation}
 m_b({\rm pole}) = 4.82\pm 0.03\; {\rm GeV}.
\label{eq.3.09} 
\end{equation}
Thus all fundamental quantities used in our calculations: $m_b$ (pole),
$\Lambda_{\rm V}$ and the string tension $\sigma\approx 0.18\pm 0.02$
GeV$^2$ (taken from the slope of the Regge trajectories for mesons) are
taken  in correspondence to QCD;  they  are determined in the narrow
ranges (\ref{eq.3.06}) and (\ref{eq.3.09}).

\section{Relativistic corrections}
\label{sec.04}

Here for the $b\bar b$ spectrum  we try to calculate the splittings
$\Delta_1$ and $\Delta_2$  defined in Eq. (\ref{eq.1.01}) with high
accuracy. First of all it is of interest to compare the spin-averaged
masses $M(nL)$ in the relativistic case, where $M(nL)$ coincides  with
the eigenvalue (e.v.)  of the spinless Salpeter equation (SSE),
\begin{equation}
 \left(2 \sqrt{\vep^2+m^2} + V_{\rm st} (r) \right)\, \psi (nL) =
 M(nL) \psi(nL),
\label{eq.4.01}
\end{equation}
with the e.v. $\widetilde M(nL)$ in the nonrelativistic (NR) limit
which is widely used in bottomonium,
\begin{equation}
 \widetilde M(nL)=2m+ E(nL).
\label{eq.4.02}
\end{equation}
Here the ``excitation energy'' $E(nL)$ is the e.v. of the
Schr\"{o}dinger equation with the same static potential  as in
Eq.~(\ref{eq.4.01}):
\begin{equation}
 V_{\rm st} =\sigma r - \frac{4}{3} \frac{\alpha_{\rm B}(r)}{r}.
\label{eq.4.03}
\end{equation}
The mass $m=m_b({\rm pole})$ and $\alpha_{\rm B}(r)$ will be taken  below in 2-loop
approximation.

In Section~\ref{sec.05} we also present the $b\bar b$ spectrum calculated
with  the Cornell potential which has been successfully used for many
years \cite{ref.08}:
\begin{equation}
 V_C(r) =\sigma r -\frac43
\frac{\alpha_{\rm static}}{r}+C_0,
\label{eq.4.04}
\end{equation}
with $\alpha_{\rm static}=const$. Since in $\alpha_{\rm static}$ the
(AF) behavior is neglected, this coupling can be considered as an
``effective freezing'' coupling in the gluon-exchange potential. Notice
that $\alpha_{\rm B}(r)$ in BPT appears indeed to be close to a
constant (the critical value) already at rather small $Q\bar Q$
separations, $r\ga 0.4$ fm;  e.g. for $n_{\rm f}=5$ the ratio
$\alpha_{\rm B}(r) /\alpha_{\rm crit}$ is equal to 0.83, 0.88, 0.92,
and 0.95 respectively, for $r$ equal 0.4 fm , 0.5 fm, 0.6 fm, and 0.7
fm.

In Table~\ref{tab.01} the spin-averaged $b\bar b$ masses  below the
$B\bar B$  threshold, the splittings
\begin{equation}
 \Delta_1= \Upsilon (1D)-\Upsilon({\rm 1P}),\quad
 \Delta_2 = \Upsilon({\rm 2P}) - \Upsilon({\rm 1P}), \quad
 \Delta_3 =\Upsilon ({\rm 2D}) -\Upsilon ({\rm 2P}),
\label{eq.4.05}
\end{equation}
and the relativistic corrections $\delta_{\rm R} = M(nL)- \widetilde{M}(nL)$
are presented for the static potential in BPT Eq.~(\ref{eq.4.03}) with a
typical set of parameters taken from the ranges (\ref{eq.3.06}) and
(\ref{eq.3.09}):
\begin{equation}
 m=4.828\; {\rm GeV},\quad \sigma =0.178\;{\rm GeV}^2,\quad
 \Lambda_{\rm V}^{(5)} =330\; {\rm MeV},\quad M_{\rm B}=1.0\; {\rm GeV}.
\label{eq.4.06}
\end{equation}

The accuracy of our numerical calculations is $\pm 1$ MeV for the e.v.
of the SSE and $\pm 0.5$ MeV for the Schr\"{o}dinger equation.

\begin{table}[ht]
\caption {\label{tab.01} The  spin-averaged $b\bar b$ masses $M(nL)$
for the SSE (\ref{eq.4.01}), $\widetilde{M}(nL)$ (\ref{eq.4.02}) in the
NR case, the splittings $\Delta_1$, $\Delta_2$, and $\Delta_3$
(\ref{eq.4.05}), and the relativistic corrections $\delta_{\rm R}$ (all
in MeV) for the static potential in BPT (\ref{eq.3.01}) with the
parameters (\ref{eq.4.06}).}
\begin{center}
\begin{tabular}{|r|r|r|r|}
\hline
     &  $M(nL)$ &   $\widetilde M(nL)$&\\
State&  SSE~~  & NR~~ & $\delta_{\rm R}$\\ 
\hline
 1S~ & 9470& 9487&-17\\
 2S~ &  10022& 10042& -20\\
 3S~ & 10368 & 10393& -24\\
 1P~ & 9900& 9909& -9\\
 2P~ & 10266&10282&-16\\
 3P~ &10555 & 10577 &-22\\
 1D~ & 10157& 10166& -9\\
 2D~ & 10458 & 10474&-16\\
 $\Delta_1$~ & 257&257& 0\\
 $\Delta_2$~ & 367& 374&-7\\
 $\Delta_3$~ &191&191&0\\
\hline
\end{tabular}
\end{center}
\end{table}

From Table~\ref{tab.01} one can see that the relativistic corrections
$\delta_{\rm R}$ give a contribution ($\sim 10 - 22$ MeV) to the absolute
values of the $b\bar b$ mass, nevertheless the splittings $\Delta_1$
and $\Delta_3=M$(2D)$-M$(2P) for the SSE and in the NR case appear to
be equal (within $\pm$ 1 MeV). This result is practically independent
of the choice of such parameters as $\sigma$ and $m$. Therefore
$\Delta_1$ can be used as an important criterion to distinguish between
different choices of the strong coupling. It can be shown that for the
S-wave states $\delta_{\rm R}$ is larger and turns out to be more
sensitive to the choice of $m$ and $\sigma$.

In the splitting $\Delta_2 =M({\rm 2P}) -M({\rm 1P})$ the relativistic
corrections are partly cancelled with $\Delta_2^{\rm NR}$  being 7 MeV
larger than $\Delta_2$ for the SSE. Therefore it is better to use the
SSE for the calculation of the highly excited $b\bar{b}$ states.  (For
the Cornell potential we find practically the same values for
$\delta_{\rm R}$.)

\section{Analysis of the spectrum}
\label{sec.05}

To demonstrate the sensitivity of the $b\bar b$ masses and the
splittings to the choice of $\Lambda_{\rm V}^{(n_{\rm f})}$ present in
the coupling $\alpha_{\rm B}(r)$ we give in Table~\ref{tab.02} the
e.v.s  $M(nL)$ of the SSE for three different values of $\Lambda_{\rm
V}^{(5)}$ in two-loop approximation taken from the range
(\ref{eq.3.06}) $(\sigma =0.178$ GeV$^2$, $M_{\rm B}=0.95$ GeV). We
denote these three sets as
\begin{eqnarray}
 {\rm I} & &  m=4.819\;{\rm GeV}\,\quad
 \Lambda^{(5)}_{\rm V}=300\;{\rm MeV},\quad
 \Lambda_{\overline{\rm MS}}^{(5)}=200\;{\rm MeV},
\nonumber \\
 {\rm II} & &  m=4.830\;{\rm GeV}\,\quad
 \Lambda^{(5)}_{\rm V}=320\;{\rm MeV},\quad
 \Lambda_{\overline{\rm MS}}^{(5)}=234\;{\rm MeV},
\nonumber \\
 {\rm III} & &  m=4.836\;{\rm GeV}\,\quad
 \Lambda^{(5)}_{\rm V}=340\;{\rm MeV},\quad
 \Lambda_{\overline{\rm MS}}^{(5)}=249\;{\rm MeV}.
\label{eq.5.01}
\end{eqnarray}

\begin{table}[ht]
\caption {\label{tab.02} The $b\bar b$ spin-averaged masses $M(nL)$ (in
MeV) for the SSE and the splittings $\Delta_1=M({\rm 1D})-M({\rm 1P})$,
$\Delta_2= M({\rm 2P})-M({\rm 1P}) $, and  $\Delta_3 =M({\rm
2D})-M({\rm 2P})$ for different values of $m$ and $\Lambda_{\rm V}^{(5)}$ (or
$\Lambda_{\overline{\rm MS}}^{(5)}$) given by Eq.~(\ref{eq.5.01}). The
values $\sigma=0.178$ GeV${}^2$ and $M_{\rm B}=0.95$ GeV were kept fixed.}
\begin{center}
\begin{tabular}{|r|r|r|r|l|}
\hline
State & I~~~ & II~~ & III~ & ~experiment\\
\hline
 1S~ &   9484& 9473&9468& $~9460.3 \pm 0.3$ ($1^3S_1$)\\
 2S~ &  10023& 10023& 10023& ~$10023.3\pm 0.3$ ($2^3S_1$)\\
 3S~ & 10364 & 10370& 10372& ~$10355.2\pm 0.5$ ($3^3S_1$)\\
 1P~ & 9900& 9900& 9900& ~$9900.1\pm 0.6$ \\
 2P~ & 10262&10266&10269& ~$10260.0\pm 0.6$ \\
 1D~ &10152& 10157&10159& ~$10162.2\pm 1.6$ (exp)${}^{+1.0}_{-0.0}$ (th)\\
 2D~ & 10450 & 10457& 10461& ~$-$ \\
 $\Delta_1$~ & 252&257& 259& ~$262.1 \pm 2.2~(\exp)^{+1.0}_{-0.0}$ (th)\\
 $\Delta_2$~ & 362& 366& 369& ~$360.1 \pm 1.2$ \\
 $\Delta_3$~ &188&190&192& ~$-$ \\ 
\hline
\end{tabular}
\end{center}
\end{table}

From Table \ref{tab.02} one can see that the splitting $\Delta_1$
increases with growing $\Lambda^{(5)}_{\rm V}$ (or
$\Lambda_{\overline{\rm MS}}^{(5)}$) and reaches the experimental value
at the value $\Lambda^{(5)}_{\rm V}\approx 325\pm 10$ MeV which
corresponds to
\begin{equation}
 \Lambda_{\overline{\rm MS}}^{(5)} (2-{\rm loop} ) \approx 238\;\pm 7\;{\rm MeV},
 \quad \alpha_s(m_Z) =0.1189\pm 0.0005
\label{eq.5.02}
\end{equation}
and gives rise to the critical value ($M_{\rm B}=0.95$ GeV)
\begin{equation}
 \alpha_{\rm crit} (n_{\rm f}=5) \approx 0.58.
\label{eq.5.03}
\end{equation}
Thus, the conventional $\Lambda^{(5)}_{\overline{\rm MS}}$ (\ref{eq.5.02})
(close to the upper limit (\ref{eq.2.07})) appears to be at the same time
compatible with the large freezing value of the background
coupling, giving rise to a good description of the $b\bar b$ spectrum
and of the splittings $\Delta_1$ and $\Delta_2$. Note that the
critical value (\ref{eq.5.03}) is in striking agreement with
$\alpha_{\rm crit} =0.60$ introduced in Ref. \cite{ref.16} in a
phenomenological way.

This important statement does depend neither on the value chosen for
$\sigma$ nor on the quark mass. To illustrate this fact we give in
Tables \ref{tab.03} and \ref{tab.04} the splittings $\Delta_1$ and
$\Delta_2$ as well as the 1P-1S and 2S-1S splittings for different
values of $\sigma$ and two values of $\Lambda_{\rm V}(n_{\rm f} = 5)$:
$\Lambda_{\rm V}^{(5)}= 280$ MeV which corresponds to
$\Lambda_{\overline{\rm MS}}^{(5)}=205$ MeV and $\Lambda_{\rm V}^{(5)}=335$
MeV which corresponds to $\Lambda_{\overline{\rm MS}}^{(5)}= 245$ MeV
(all $\Lambda$'s in two-loop approximation).

Below we  would like also to show the dependence of the splittings on
the value of the string tension. We take $\sigma$ in the range
0.165-0.185 GeV${}^2$ and fix the QCD constant $\Lambda^{(5)}_{\rm V}=280$
MeV. (This value of $\Lambda^{(5)}_{\rm V}$ corresponds to
$\Lambda_{\rm\overline{\rm MS}} =205$ MeV and $\alpha(M_Z)= 0.116$ for
$M_{\rm B}=1.0$ GeV.) Note that for such a value of $\Lambda^{(5)}_{\rm V}$ the
freezing value, $\alpha_{\rm crit} = 0.488$, appears to be 20\% smaller
than that in Eq.~(\ref{eq.5.03}).

From the numbers presented in Table~\ref{tab.03} one can see that the
splitting $\Delta_1$ is slightly increasing  from the value 241 MeV up
to 251 MeV while $\sigma$ is changing in a wide range, from 0.17
GeV${}^2$ to 0.185 GeV${}^2$, being still $10 - 20$ MeV smaller than
the experimental number given in Eq.~(\ref{eq.1.01}). Also other
splittings between low-lying states, e.g., $M$(2S)$ -M$(1S) and
$M$(1P)$-M$(1S), appear to be by $30-50$ MeV smaller than the
experimental numbers for any $\sigma$ from the range $0.17-0.185$
GeV${}^2$.

Our calculations show that for a reasonable value of $\Lambda^{(5)}_{\rm
V}$ one cannot reach agreement with experiment by variation of the
parameter $\sigma$ only. Larger values of $\Lambda^{(5)}_{\rm V}$ (around 320
MeV) are needed to get $\Delta_1 \approx 260$ MeV.

\begin{table}[ht]
\caption {\label{tab.03} The splittings $\Delta_1= M$(1D)$-M$(1P),
$\Delta_2= M$(2P)$ - M$(1P), $M$(2S)$ - M$(1S), and $M$(1P)$ - M$(1S)
(in MeV) for $\Lambda_{\rm V}^{(5)}= 280$ MeV ($\Lambda_{\overline{\rm
MS}}^{(5)}= 205$ MeV) and different $\sigma$ (in GeV${}^2$) ($M_{\rm B} =
1.0$ GeV, $m_b=4.82$ GeV).}
\begin{center} 
\begin{tabular}{|l|r|r|r|r|r|l|} \hline
 ~State~ & ~$\sigma=0.170$~ & ~$\sigma=0.174$~ & ~$\sigma=0.178$~ 
& ~$\sigma=0.182$~  & ~$\sigma=0.185$~  & ~experiment \\
\hline
 ~$\Delta_1$~ & 241~ & 243~ & 246~ & 251~ & 251~ & ~$262.1 \pm 2.2$  \\
 ~$\Delta_2$~ & 347~ & 351~ & 356~ & 360~ & 363~ & ~$360.1 \pm 1.2$  \\
 ~$M$(1P)$ - M$(1S)~ & 396~ & 399~ & 402~ & 406~ &~ 408 & $\approx 430^a$  \\
 ~$M$(2S)$ - M$(1S)~ & 514~ & 518~ & 523~ & 529~ & ~533 & $\approx 550^a$  \\
\hline 
\end{tabular} 
\end{center}
\end{table}

The situation appears to be different if a large value
$\Lambda_{\overline{\rm MS}}^{(5)}(2-{\rm loop})=245$ MeV ($\alpha(M_Z)=0.1194$)
is taken ( see Table \ref{tab.04}). In this case the same three
splittings turn out to be in very  good agreement with the experimental
numbers for $\sigma=0.18 \pm 0.02$ GeV${}^2$.

\begin{table}[ht]
\caption {\label{tab.04} The same splittings as in Table \ref{tab.03} for 
$\Lambda_{\rm V}^{(5)}=335$ MeV 
($\Lambda_{\overline{\rm MS}}^{(5)}(2-{\rm loop})=245$ MeV)}
\begin{center}
\begin{tabular}{|l|r|r|r|r|r|l|} \hline
 ~State~ & ~$\sigma=0.170$~ & ~$\sigma=0.174$~ & ~$\sigma=0.178$~ 
& ~$\sigma=0.182$~  & ~$\sigma=0.185$~  & ~experiment \\
\hline
 ~$\Delta_1$~ & 253~ & 255~ & 258~ & 261~ & 263~ & ~$262.1 \pm 2.2$  \\
 ~$\Delta_2$~ & 360~ & 364~ & 368~ & 372~ & 376~ & ~$360.1 \pm 1.2$  \\
 ~$M$(1P)$ - M$(1S)~ & 423~ & 426~ & 430~ & 433~ &~ 436 & $\approx 430^a$  \\
 ~$M$(2S)$ - M$(1S)~ & 544~ & 549~ & 554~ & 559~ & ~563 & $\approx 550^a$  \\
\hline
\end{tabular}
\end{center}
\end{table}

While the calculated 1D-1P, 2S-1S, and 1P-1S splittings are in precise
agreement with experiment for $\Lambda_{\rm V}^{(5)}=320-335$ MeV
($\Lambda_{\overline{\rm MS}}^{(5)} = 234-245$ MeV), the 2P-1P
splitting in these cases appears to be $\approx 10$ MeV larger than the
experimental one. We expect that a small decrease of $M$(2P) (and also
of $M$(3S)) can be reached taking into account  the flattening of the
confining potential due to additional pair creation.

The sensitivity of the $b\bar b$ spectrum to the Coulomb constant in
the Cornell potential is illustrated in Table~\ref{tab.05} where three
variants with fixed  $m=4.80$ GeV and $\sigma=0.179$ GeV$^2$ are presented
while $\alpha_{\rm static} = constant$ is varied.  From
Table~\ref{tab.05} one can see that only for the choice with large
Coulomb constant $\alpha_{\rm static} = 0.41 - 0.43$, good agreement
with experiment is obtained for all states with the exception of the
ground state and partly of the 2S state.

\begin{table}[h]
\caption {\label{tab.05} The  $b\bar b$ spin-averaged masses
$\tilde M(nL)$ (NR case) for the Cornell potential ($m=4.80$ GeV,
$\sigma=0.179$ GeV${}^2$) with different values of $\alpha_{\rm static}$
and $C_0$ (the masses and $C_0$ are given in MeV)}
\begin{center}
\begin{tabular}{|l|r|r|r|l|}
\hline
 & $\alpha_{\rm st}=0.3645$& $\alpha_{\rm st}=04071$&
 $\alpha_{\rm st}=0.4263$&\\
 & ~~$C_0=-74$ ~ & ~~$C_0=-32$ ~ & ~~$C_0=-16$ ~ & ~~experiment\\
\hline
 $\widetilde{M}({\rm 1S}) +\delta_{\rm AF}({\rm 1S})^{a)}$
 & 9421 ~ & 9434 ~ & 9451 ~ & ~~$9460.3 \pm 0.3 ~(1{}^3S_1)$ \\
 $\delta_{\rm AF}$ (1S) & -34 ~ &  +37 ~ & +79~ &\\
 $ 2S^{b)}$& 10000~ & 9995~ & 9993~ & ~~$10023.3 \pm 0.3 ~(2^3S_1)$\\
 $ 3S$ &10337~ &10344~ & 10347~ & ~~$10355.2 \pm 0.5 ~(^3S_1)$\\
 1P & 9900 (fit) ~ & 9900 (fit) ~ & 9900(fit) ~ & ~~$9900.1\pm 0.6$\\
 2P & 10249 ~ & 10259 ~ & 10263 ~ & ~~$10260.0 \pm 0.6$\\
 1D & 10139 ~ & 10153 ~ & 10158 ~ & ~~$10162.2 \pm 1.6^{+ 1.0}_{-0.0}$\\
 2D & 10430 ~ & 10448 ~ & 10458 ~ & ~~- \\
 $\Delta_1$ & 239~ &253 ~ & 258 ~ & ~~$262.1\pm 2.2({\rm exp})^{+1.0}_{0.0}$(th)\\
 $\Delta_2$ & 347 ~ & 359 ~ & 363 ~ & ~~$360.1 \pm 1.2$\\
 $\Delta_3$ & 183 ~ & 189 ~ & 192.5 ~ & ~~- \\ 
\hline
\end{tabular}
\end{center}

$^{a)}$ The AF correction $\delta_{\rm AF}$(1S) is defined in
Eq.~(\ref{eq.5.04})

$^{b)}$ The AF correction is neglected for all states with the exception of
the ground state.

\end{table}
For the ground state $\Upsilon$(1S) the AF behavior is very important
both for the mass and for the wave function at the origin. Therefore
one may take into account the difference between $\alpha_{\rm static}$
in the Cornell potential (playing the role of the effective freezing
coupling) and $\alpha_{\rm B}(r)$ considering it as a perturbation.  
The perturbative treatment gives rise to the mass correction
\begin{equation}
 \delta_{AF} ({\rm 1S})=\frac{4}{3} \left\langle \frac{\alpha_{\rm
 static} -\alpha_{\rm B}(r)}{r} \right\rangle_{1{\rm S}}.
\label{eq.5.04} 
\end{equation} 
Owing to the AF effect this correction turns out to be positive (see
second row in Table~\ref{tab.05}) for $\alpha_{\rm static} \ga 0.39$.
However, a relatively large value for the AF correction, $\delta_{\rm
AF} \approx 40 -80$ MeV, for the ground state clearly indicates that
the AF effect in the gluon-exchange potential is very important and
cannot be treated accurately as a perturbation to the Coulomb potential.

The meaning of the quantity $\alpha_{\rm static}$ in the Cornell
potential can be understood if one introduces in BPT the effective
coupling for a given $nL$ state according to the following relation:
\begin{equation}
 \left\langle \frac{\alpha_{\rm B}(r)}{r} \right\rangle_{nL} =
 \alpha_{\rm eff} (nL) \left\langle \frac{1}{r}\right\rangle.
\label{eq.5.05}
\end{equation}

\begin{table}[ht]
\caption {\label{tab.06}  The   effective coupling $\alpha_{\rm
eff}(nL)$~(\ref{eq.5.05}) for the static potential with the
parameters (\ref{eq.4.06})}
\begin{center}
\begin{tabular}{|l|lllllll|}
\hline
 ~state~& ~1S~&~2S~&~3S~&~1P~&~2P~&~1D~&~2D\\ 
\hline
 $\alpha_{\rm eff}(nL)$~&~0.39~&~0.41~&~0.43~&~0.47~&~0.47~&~0.50~&~0.50\\ 
\hline
\end{tabular}
\end{center}
\end{table}
The values of $\alpha_{\rm eff} (nL)$ are given in Table~\ref{tab.06} from
which one can see that for the low-lying states the effective coupling in
BPT, $\alpha_{\rm eff} \approx 0.40 - 0.43$ is indeed very close to the
Coulomb constant used in the Cornell potential.

From the spectra presented in Table~\ref{tab.05} one can conclude that the
splittings $\Delta_1$ and $\Delta_2$ are in precise agreement with
experiment for 
\begin{equation}
 \alpha_{\rm static} =0.43\pm 0.02
\label{eq.5.06}
\end{equation}
and this value appears to be 25\% lower than the critical value for the
background coupling states given in Eq.~(\ref{eq.5.03}) and very close
to the effective coupling given in Table~\ref{tab.06}.

For the best variants from Tables \ref{tab.02} and \ref{tab.05} the
splitting $\Delta_3 = M(2{\rm D}) - M$(2P) is practically unchanged,
$\Delta_3= 191\pm2$ MeV and therefore our prediction for the mass of
the excited 2D state is $ M_{\rm cog} ({\rm 2D}) \approx M({\rm 2D}_2)=
10451\pm 2$ MeV.

\section{Conclusion}
\label{sec.06}

From our analysis of the bottomonium spectrum we observe that the
splittings between orbital excitations $\Delta_1$, $\Delta_2$, and
$\Delta_3$ appear to be very sensitive to the freezing value of the
coupling in the gluon-exchange potential.  The splitting $\Delta_1=
M$(1D)$-M$(1P) is of special importance since it is measured with the
great accuracy

For the Cornell potential, when the Coulomb constant
$\alpha_{\rm static}$ can be considered as an effective freezing
coupling, precise agreement with experiment takes place for
$\alpha_{\rm static}=0.43\pm 0.02$.

For the $Q\bar Q$ gluon-exchange potential with the strong
coupling taken as in BPT, where the freezing value $\alpha_{\rm crit}$
is fully determined by the value of $\Lambda_{\rm V}(n_{\rm f}=5)$ in the Vector-scheme, good
agreement with experiment is obtained only for
$\Lambda_{\rm V}^{(5)}= 325\pm 10$ MeV which corresponds to the
conventional $\Lambda_{\overline{\rm MS}}^{(5)}$ close to the upper
limit: $\Lambda_{\overline{\rm MS}}^{(5)}$ (2-loop)= 238$\pm 7$ MeV
and $\alpha_s(M_Z)=0.119\pm0.001$ and at the same time gives rise
to a large freezing value, $\alpha_{\rm crit} \approx 0.58$.

We also predict that the mass of the as yet unobserved 2D state(s) is
$M(2{\rm D}_J)\approx M_{\rm cog} ({\rm 2D}) =10451\pm 2$ MeV.

\end{document}